# Influence of interlayer exchange coupling on ultrafast laser-induced magnetization reversal in ferromagnetic spin valves

Junta Igarashi[1,*], Yann Le Guen[1], Julius Hohlfeld[1], Stéphane Mangin[1,2], Jon Gorchon[1], Michel Hehn[1,2], and Grégory Malinowski[1]

[1] Université de Lorraine, CNRS, IJL, F-54000 Nancy, France
[2] Center for Science and Innovation in Spintronics, Tohoku University, 2-1-1 Katahira, Aoba-ku, Sendai 980-8577 Japan

[*] junta.igarashi@univ-lorraine.fr, jigarashi1993@gmail.com

## ABSTRACT

In this study, we explore the influence of interlayer exchange coupling on magnetization reversal triggered by femtosecond laser pulses in ferromagnetic spin valves. Our experiments, focused on femtosecond laser-induced magnetization reversal, methodically vary the thickness of the copper (Cu) spacer layer. We identify a critical Cu thickness threshold at 2.32 nm. Above this threshold, a stable reversed magnetic domain is consistently generated upon exposure to a single laser pulse. Conversely, with a Cu spacer thinner than 2.32 nm, the observed magnetization reversal from parallel (P) to anti-parallel (AP) states occurs only under continuous laser irradiation. Once the laser is stopped, the magnetic configuration relaxes back to its initial P state, influenced by ferromagnetic exchange coupling. This research enhances our understanding of the mechanisms that drive optically induced ultrafast magnetization reversal in ferromagnetic spin valves.

# I. INTRODUCTION

Magnetization manipulation without applied external magnetic fields is attractive from both fundamental and technological viewpoints. Current- and light- induced magnetization manipulation, e.g., spin-transfer torque (STT) [1,2] or Spin-orbit torque (SOT) [3] and all-optical switching (AOS) [4], have been developed owing to the possibility of growing thin film heterostructures. Combining both currents and ultrafast laser excitation pushes spintronics to the sub-picosecond timescale. So far, only a few studies have been performed: magnetic tunnel junction (MTJ) employing all-optical switchable materials (GdFeCo [5,6], Gd/Co bilayer [7], [Co/Tb] multilayers [8–10]) [8,11–14] and current-induced domain wall motion [15–17]. In the former case, to improve tunnel magnetoresistance ratio, a composite free layer of CoFeB exchange-coupled with AOS material is used [8,12–14]. As reported in Gd(Fe)Co coupled with Co/Pt due to Ruderman–Kittel–Kasuya–Yosida (RKKY) interaction, when magnetization reversal of Gd(Fe)Co occurs, magnetization of [Co/Pt] is also switched within 10 ps owing to the exchange coupling [18,19]. This approach has allowed ultrafast control of a variety of magnetic materials.

Recently, single-shot magnetization reversal in ferromagnetic spin valves has been demonstrated [20], which are also used for current-induced STT swicthing [21]. The direction of magnetization reversal is dependent on the fluence of the laser pulse. At lower fluences, the reversal is from parallel (P) to anti-parallel (AP) configuration, while at higher fluence, the reversal is from AP to P configuration. The magnetization reversal from the P state differs significantly from the current understanding of AOS and requires a thorough understanding of the underlying mechanism. Previous studies have indicated that magnetization reversal is caused by ultrafast spin currents generated from the ultrafast demagnetization of the free layer [20,22]. The hypothesis is that these spin currents propagate through the Cu layer to the reference layer before experiencing a spin-flip when they are reflected at the interface. The reflected spin currents are then injected into the free layer, causing the magnetization in the free layer to recover in the opposite direction. This mechanism can be understood by analogy with current-induced STT [1,2]. Numerical calculations, which consider the reflection mechanism phenomenologically, could reproduce experimental results qualitatively [22]. However, the physical mechanism leading to the spin-flip upon reflection and the magnetization reversal has not been fully understood yet.

In a previous study, the spin valve structure was made of two [Co/Pt] multilayers separated by a Cu spacer with a 5 nm or larger thickness. The interlayer exchange coupling between the two ferromagnetic layers was negligibly small. In this paper, we study whether interlayer exchange coupling can assist ultrafast spin currents-induced magnetization reversal, as in ferrimagnetic/ferromagnetic systems [18,19]. We investigate the evolution of interlayer exchange coupling and magnetization reversal induced by femtosecond laser pulses in fully ferromagnetic spin valves as a function of the Cu thickness. Using a Cu wedge spin valve, a critical Cu thickness at which a stable reversed domain can be defined.

## II. EXPERIMENT

### A. Sample structure

The prepared spin valves consist of Glass (substrate) /Ta(5)/Pt(4)/[Co(0.82)/Pt(1)]$_3$/Co(0.82)/Cu($t_{Cu}$)/[Co(0.49)/Pt(1)]$_2$/Ta(5) prepared by magnetron sputtering (Fig. 1(a)). The numbers in parentheses represent the nominal thickness of each layer in nm. Using the wedge deposition method, the thickness of Cu linearly varies from 1.00 to 3.00 nm along a 50 mm substrate (see Ref. [10] for details about wedged deposition). Here, we define bottom and top [Co/Pt] multilayers as the reference and free layers, respectively.

Hysteresis loops obtained by magnet-optic Kerr effect (MOKE) with an applied magnetic field perpendicular to the film plane for various Cu thicknesses are shown in Figure 1(b). For $t_{Cu}$ ranging from 1.00 nm to 2.20 nm, the free and reference layers are strongly ferromagnetically coupled, which results in a single-step magnetization reversal. In Supplementary Material Note 1 [23], we show MOKE hysteresis loops obtained by applying an external field up to 700 mT. For $t_{Cu}$ > 2.20 nm, four magnetic configurations with distinct switching fields ($H_{SW1}$ and $H_{SW2}$) are observed, as reported in a previous study [24]. In Fig. 1(c), we summarize the evolution of the coercive field $H_C$ and the shift field $H_{Shift}$ extracted from the minor hysteresis loops, which are defined as $H_C$ ($H_{Shift}$) = ($H_{SW1}$- (+) $H_{SW2}$)/2. $H_{SW1}$ and $H_{SW2}$ are the switching fields indicated in Fig. 1(b). For 2.20 nm < $t_{Cu}$ < 2.60 nm, $H_{shift}$ is negative, which means that both layers are ferromagnetically coupled [25]. For $t_{Cu}$ > 2.60 nm, we can see a slight positive shift in the minor loops, meaning that the two ferromagnetic layers are weakly antiferromagnetically coupled. This oscillation of the exchange coupling with the spacer thickness is a well-known signature of RKKY coupling [26].

## B. Single-shot AOS

Here, we present magnetization reversal induced by a single linearly polarized femtosecond-laser pulse, as reported in a previous study [20]. A Ti: sapphire femtosecond-laser source and regenerative amplifier for the pump laser beam were used to study single-shot switching measurement. Wavelength, pulse duration, and repetition rate of the fs laser were 800 nm, ≈ 50 fs, and 5 kHz, respectively. The $1/e$ spot size $2w_0$ is ≈ 70 μm. The femtosecond-laser pulses were shined from the free layer side, as indicated in Fig. 2(a). MOKE imaging was performed from the other side of the sample using an LED source with a wavelength of 628 nm. We normalized MOKE images with the difference in the extracted gray level between P and AP state. Since we couldn't directly confirm the AP state in thinner Cu, we defined the gray level between P and AP states in the thinner Cu region using a linear fit as a function of the Cu thickness (see Fig. 2(a)).

Figure 2(a) shows typical images obtained after exciting the sample with a single laser pulse. For $t_{Cu}$ = 1.00, 2.12, and 2.20 nm, no switching is observed. We present only the results obtained starting from the P state (red) since access to the AP state was restricted by the strong ferromagnetic coupling. For $t_{Cu} \geq 2.32$ nm, a clear P (red) -to-AP (blue) switching induced by a single-shot laser pulse is observed. AP-to-P switching is also demonstrated for these Cu thicknesses but at higher fluences. Figure 2(b) shows MOKE images and their line profiles for $t_{Cu}$ = 2.28 nm. Although single-shot switching is observed, domains in the AP state shrink and disappear within a few seconds due to domain wall motion triggered by ferromagnetic coupling. This phenomenon was also suggested in a previous study on AOS for Gd/Co [27] (shown in Fig. 5 from IV to VI in Ref. [27]).
Here, a critical Cu spacer thickness of 2.32 nm, which enables stable P-to-AP switching with a single-shot laser pulse, can be defined. The same experiment with the laser shining on the reference layer is presented in Supplementary Material Note 3, and a similar trend is observed.

In Fig. 3, The evolution of the threshold fluences ($F_{th}$) for the P-to-AP switching $F_{th\_PAP}$ (black symbols), AP-to-P switching $F_{th\_APP}$ (red symbols), and multidomain formation $F_{th\_multi}$ for both layers (blue symbols) are summarized. $F_{th}$ is extracted for each initial state by analyzing the MOKE images for various laser pulse energies (see Supplementary Material Note 2 for details about the analysis [23]. The same formula was used for fitting, as described in Ref. [28]) as a function of the Cu spacer thickness, as shown in Fig. 3(a). $F_{th\_PAP}$ and $F_{th\_APP}$ decrease with increasing $t_{Cu}$. $F_{th\_multi}$ increases up to $t_{Cu}$ = 2.28 nm and then begins to decrease. Regarding $F_{th\_multi}$ at $t_{Cu}$ < 2.28 nm, the increase in $F_{th\_multi}$ with $t_{Cu}$ can be explained by the fact that a thicker Cu layer results in less energy absorption in the reference layer when the laser is incident on the free layer [20]. One possible reason for the decrease in $F_{th}$ with increasing $t_{Cu}$ for $t_{Cu}$ > 2.28 nm is the overestimation of $F_{th}$ due to the exchange coupling at the thinner Cu spacer. As shown in Fig. 2(b), strong ferromagnetic coupling might cause a reduction in the size of domains in the AP state. Since we extract $F_{th}$ by analyzing domains in MOKE images taken a few seconds after the laser excitation, the actual domain size might be larger than the one we extract.

To cover a broader range of thickness dependence, we measured samples used in a previous study and added data in Fig. 3(b) [20]. The $F_{th}$ for all states increases in the range above 5 nm Cu spacer. If $F_{th}$ has not been overestimated for thinner $t_{Cu}$ there might be a region between 3-5 nm, where magnetization reversal can be achieved with even less energy. We also observed the same trend with the laser shining on the reference layer (see Supplementary Material Note 3 [23]).

### C. All-optical switching with multiple laser pulses

From the single-shot magnetization reversal experiment, we determined a critical Cu spacer thickness of 2.32 nm, above which "stable" P-to-AP switching is observed. However, this does not imply that the P-to-AP switching mechanism is no longer possible below this critical Cu thickness. As demonstrated in Fig. 2(b) for $t_{Cu}$ = 2.28 nm, P-to-AP switching is still observed immediately after the laser pulse excitation, but the reversed domains disappear in the following seconds. This suggests the existence of two dynamics that will influence the final state, possibly governed by two different mechanisms. Consequently, a single-shot experiment and an observation of the final state a long time after the pulse will allow for a proper study and separation of these underlying physical mechanisms. To this aim, multiple pulses with different repetition frequencies $f$ were applied to the samples, as shown in Fig. 4(a). MOKE images obtained during continuous excitation (indicated as Laser ON in Fig. 4(a)) and after the constant excitation (indicated as Laser OFF in Fig. 4(a)) for $t_{Cu}$ = 2.16, 2.20, and 2.24 nm are presented in cases where stable reversed domains were not observed in the single-shot experiment. It is noteworthy that these MOKE images represent averages in time over the entire process and show the frequency of magnetization reversals. As stated earlier, a gray level between the P and AP states in the low Cu thickness range had to be assumed because of the strong ferromagnetic coupling. Therefore, we estimated the gray level with 95% confidence interval using a linear fit as a function of the Cu thickness (see Fig. 2(a)). At $t_{Cu}$ = 2.16 nm, the upper (lower) limit of 95% confidence interval differs by 6.5% from the estimated gray level, which is still acceptable to evaluate magnetization reversal. The laser repetition frquency $f$ is varied from 5 kHz to 5 Hz. At $t_{Cu}$ = 2.16 nm, clear P-to-AP switching is observed for $f$ = 5 kHz. However, for $f$ lower than 5 kHz, no P-to-AP switching is observed. In the case of $t_{Cu}$ = 2.20 (2.24) nm, clear P-to-AP switching is observed down to $f$ = 500 (50) Hz. However, once the laser is switched off, the reversed domains disappear in all cases.

### D. Time-resolved MOKE imaging

In a static experiment with multiple laser pulses, we have shown that the P-to-AP switching depends on the values of $f$ and $t_{Cu}$. However, static measurements do not allow us to access the transient state during the interval between laser pulses. Therefore, we perform time-resolved (TR-) MOKE imaging experiment (Fig. 5). The TR-MOKE microscope used in this study is similar to that described in Ref. [29]. The laser pulses were generated by Yb base laser (PHAROS, Light Conversion) at a wavelength of 1030 nm, then split into the pump and the probe. An optical parametric amplifier (ORPHEUS-HP, Light Conversion) was used to change the wavelength of the probe beam to 550 nm. The repetition rate of the laser was varied from 5 kHz to 100 kHz. The sample was pumped at normal incidence and probed at 45° from the free layer side. A combination of a half-wave plate and a polarizer was used to change the pump powers. The pump was focused with a 1 m lens and has a Gaussian spatial fluence distribution with a standard deviation of approximately 95 μm. At the sample, the pump and the probe Gaussian pulses width was around 200 fs. A set polarizer and analyzer were inserted in the probe path before and after the sample to obtain magnetic contrast by MOKE. The magnetic state was imaged with a 50 mm lens, and a filter was used to prevent the pump beam to reach the CMOS camera. Note that we didn't apply an external magnetic field during this experiment. In Figs. 5(a) and (b), we perform experiment with $f$ = 5 kHz. In Fig. 5(a) at $t_{Cu}$ = 2.16 nm, a domain is generated by the excitation of the laser pulse, primarily attributed to demagnetization in both layers. Given that the signal is detected in both the free and reference layers, distinguishing between demagnetization and the P-to-AP switching is challenging. However, a previous study reported that the P-to-AP switching takes place in less than a picosecond [20]. Our confirmation of the P-to-AP switching at $t_{Cu}$ = 2.16 nm in Fig. 4 suggests that not only demagnetization but also the P-to-AP switching likely occurs simultaneously in Fig. 5(a). The domain shrinks over time and disappears within 200 μs, as it is not observed at time delay $\Delta t$ = -10 ps. In contrast to the results at $t_{Cu}$ = 2.16 nm, a domain is observed at $t_{Cu}$ = 2.20 nm even at $\Delta t$ = -10 ps. We identify the domain as a reversed AP state by comparing the contrast between the obtained MOKE image and the MOKE hysteresis loop (see Supplementary Material Note 4 [23]). The reversed domain is demagnetized by laser excitation and then remains in the AP state. This indicates that the mechanism of the P-to-AP switching is independent of the initial state and always brings the domain to the AP state. Figures 5(c) and (d) show the results of TR-MOKE imaging with $f$ = 100 kHz at $t_{Cu}$ = 2.26 and 2.28 nm, respectively. At $t_{Cu}$ = 2.28 nm, we observe a reversed domain at $\Delta t$ = -10 ps, whereas it is not observed at $t_{Cu}$ = 2.26 nm. At $t_{Cu}$ = 2.26 nm, the nucleated domain due to laser excitation disappears within 10 μs. The threshold Cu thickness at $f$ = 100 kHz, where the reversed domain is observed at $\Delta t$ = -10 ps, is thicker than that at $f$ = 5 kHz.

r

### III. DISCUSSION

Here, we discuss the temporal P-to-AP switching in the thinner Cu region. We attribute the different minimum $f$ required, depending on $t_{Cu}$ (Fig. 4), to the strength of the ferromagnetic exchange coupling, which acts as an external magnetic field aligning both magnetizations to the P state. In this scenario, optical excitation transiently induces P-to-AP switching, as observed in the absence of static couplings between the layers. Unfortunately, the ferromagnetic exchange in these stacks subsequently erases the reversed AP domain. However, if another laser pulse excites the sample before the erasure of the domain, the domain will remain present and visible via static MOKE. The $f$ consequently contends with the timescale of domain erasing ($t_{ex}$), which is attributed to domain wall motion induced by exchange coupling. As the Cu thickness is reduced, the exchange coupling strength increases, and in turn, $t_{ex}$ will decrease, requiring a faster laser frequency to maintain the reversed domain. When the $t_{ex}$ is significantly shorter than $1/f$, the ferromagnetic coupling operates as an applied field that initializes the reversed domains through domain wall motion. This suggests that the magnetization reversal occurs with a frequency of $2f$, as previously predicted in the study on the AOS of the Gd/Co bilayer system in an applied magnetic field [26]. For instance, electrical detection of the magnetization reversal using current perpendicular to plane-giant magnetoresistance devices could confirm the frequency of the magnetization reversal. When $t_{ex}$ is longer than $1/f$, the reversed domain remains in the AP state despite being demagnetized after each laser pulse excitation. This implies that if spin currents are involved in magnetization reversal, the free layer always receives spin in the opposite direction to the magnetization direction of the reference layer. However, a quantitative analysis of the competition between $f$ and $t_{ex}$ using this method is challenging. This difficulty arises because MOKE images in Fig. 4 were captured during constant irradiation, providing information on the frequency of magnetization reversal. By changing the laser repetition rate, the continuous laser heating taking place in our sample may also be affected. In fact, the threshold thickness of Cu is different for $f$ = 5 kHz and 100 kHz, where the reversed domain disappears at $\Delta t$ = -10 ps (Fig. 5). According to a previous study on domain wall motion under an external magnetic field, higher temperature leads to faster domain wall velocity [30]. Thus, the continuous laser heating might elevate the temperature within the laser spot, thereby accelerating the domain wall motion and facilitating the erasure of the reversed domain.

At $t_{Cu}$ = 2.16 nm and $f$ = 5 Hz, a reversed domain is observed in Fig. 4, while it is not the case in Fig. 5(a). To understand the difference between Figs 4 and 5 at $t_{Cu}$ = 2.16 nm, we measured the radius (horizontal direction) of the reversed domains as a function of $t_{Cu}$ in Fig. 5(e). The radius of the reversed domain at $t_{Cu}$ = 2.16 nm (from Fig. 4 at $f$ = 5 kHz) for a fluence of 7.8 mJ/cm$^2$ is 7 μm shorter than that at $t_{Cu}$ = 2.24 nm. Based on Fig. 5(a), the radius of the reversed domain at $t_{Cu}$ = 2.24 nm is approximately 7 μm for a fluence of 4.5 mJ/cm$^2$. Assuming a similar domain wall velocity for both laser fluences, it can be concluded that the discrepancy between the results presented in Fig. 4 and Fig. 5(a) is due to the difference in the size of the reversed domain.

## IV. CONCLUSION

In summary, we investigate the effect of interlayer exchange coupling on magnetization reversal induced by femtosecond laser pulses in fully ferromagnetic spin valves. In contrast to the previous studies [18,19], our systematic experiments reveal that interlayer exchange coupling competes with the P-to-AP switching mechanism, where the exchange coupling acts as an effective field that reverses the written domain to reach a parallel alignment. This study contributes to a deeper understanding of the laser-induced ultrafast magnetization reversal in ferromagnetic spin valves with ultra-thin spacers.

## ACKNOWLEDGEMENTS

This work was supported by the French National Research Agency (ANR) through the France 2030 government grants EMCOM (ANR-22-PEEL-0009), UFO (ANR-20-CE09-0013), "Lorraine Université d'Excellence" (ANR-15-IDEX-04-LUE) by the Institute Carnot ICEEL for the project FASTNESS, the Région Grand Est, the Metropole Grand Nancy, by the French PIA project, the "FEDERFSE Lorraine et Massif Vosges 2014-2020", a European Union Program, the ANR project ANR-20-CE24-0003 SPOTZ, the Sakura Program, the JSPS Bilateral Program, the Tohoku University-Universite de Lorraine Matching Funds, and CSIS cooperative research project in Tohoku University. J.I. acknowledges support from JSPS Overseas Research Fellowships. All fundings were shared equally among all authors.

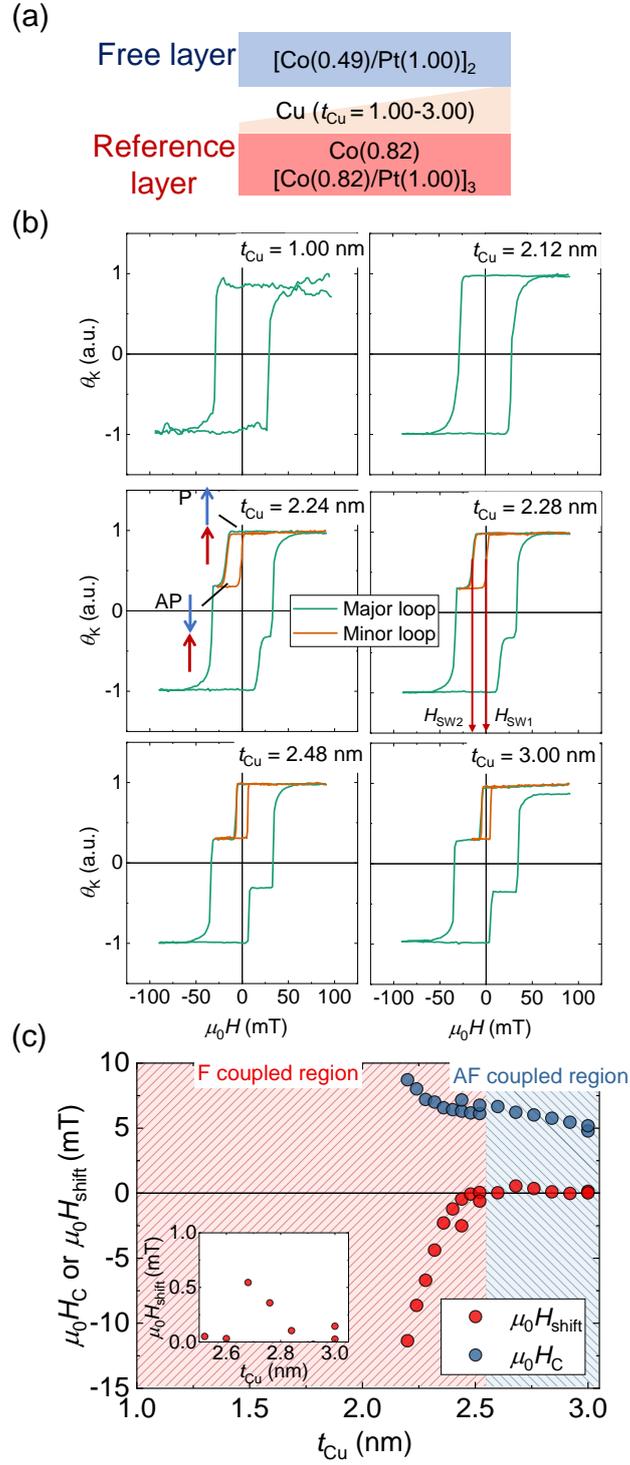

FIG. 1. Sample characterization by measuring MOKE hysteresis loops. (a) Schematic illustration of the studied sample structure with a wedge-shaped Cu spacer. (b)

Representative results of MOKE hysteresis obtained by measuring the Kerr rotation $\theta_K$ as a function of the magnetic field H applied perpendicular to the film plane . Green (orange) lines correspond to major (minor) loops. We extract switching fields $H_{SW1}$ and $H_{SW2}$ from minor loops to evaluate coercive field $H_C$ and shift field $H_{shift}$. (c) Summarized $H_C$ and $H_{shift}$ for the studied samples. The inset shows an enlarged graph from $t_{Cu}$ = 2.5 to 3.0 nm. Shaded regions indicate ferromagnetic (F) coupled and antiferromagnetic (AF) coupled regions.

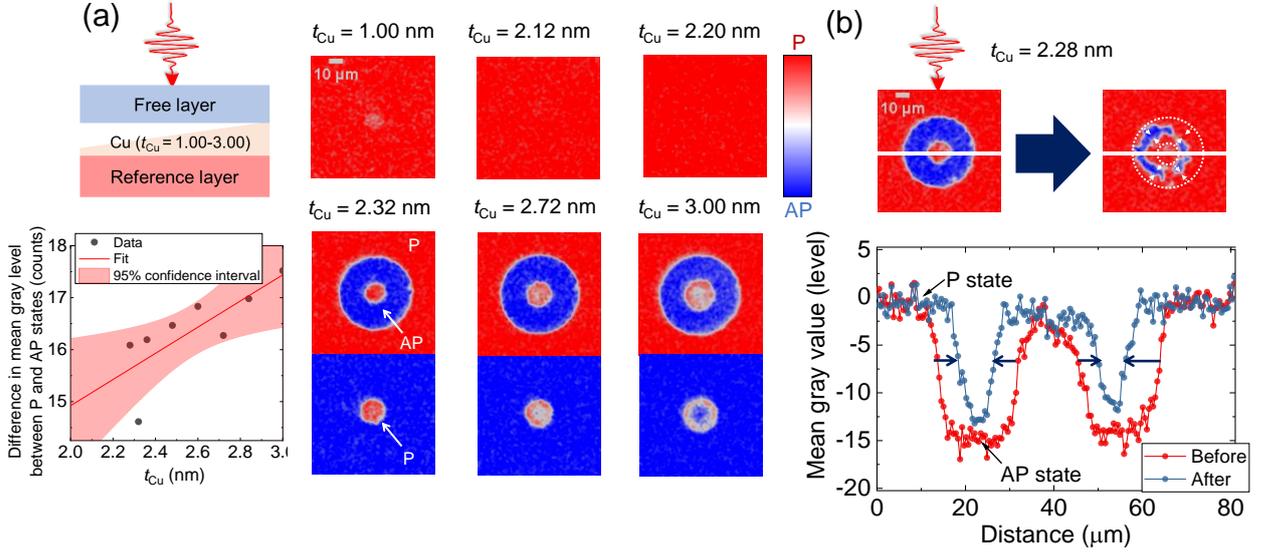

FIG. 2. Magnetization switching using a single femtosecond laser pulse excited from the free layer side. (a) MOKE images obtained after irradiation of a fs-laser pulse starting from the P state (red) and AP state (blue) for the studied samples. The laser pulse is shone from the free layer side. Difference in mean gray value between P and AP states is presented as a function of $t_{Cu}$. The line is a linear fit to the results. We also show 95% confidence interval. (b) MOKE images obtained right after a laser pulse excitation and a few seconds later for $t_{Cu}$ = 2.28 nm. Line profiles are also shown for obtained MOKE images. The laser fluence used for the experiment is around 7.8 mJ/cm$^2$.

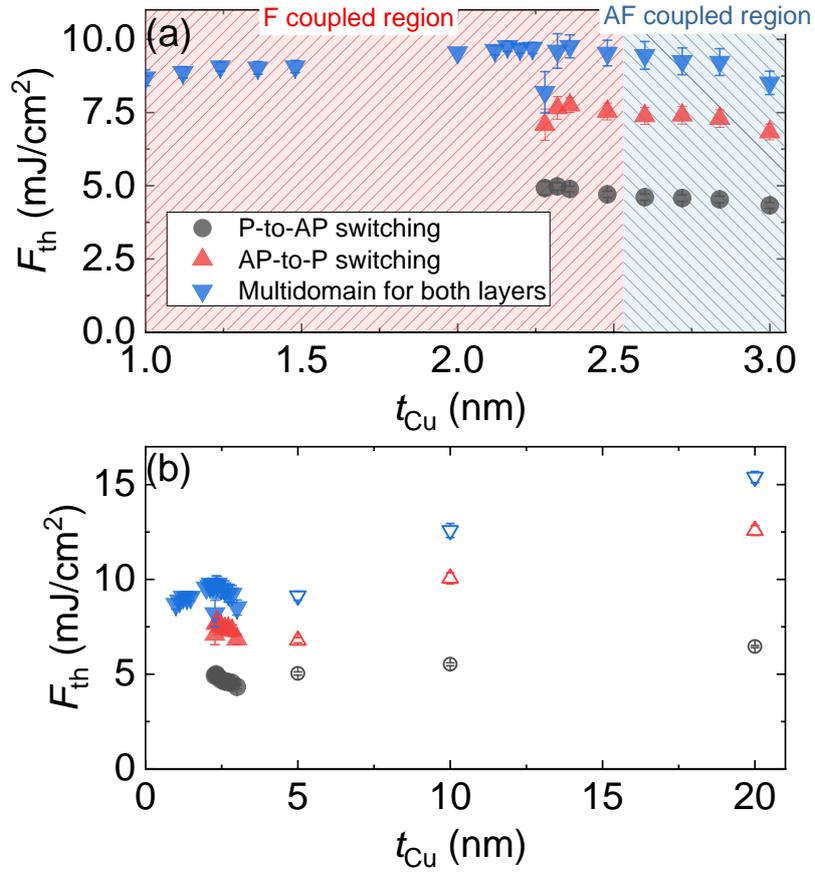

FIG. 3. Threshold fluence for P-to-AP switching (gray circles), AP-to-P switching (red triangles) and multidomain state in both layers (blue triangles) as a function of the Cu spacer thickness for (a) $t_{Cu}$ up to 3.0 nm. (b) $t_{Cu}$ up to 20 nm, including samples from a previous study (opened symbols) [20]. Shaded regions in (a) indicate ferromagnetic (F) coupled and antiferromagnetic (AF) coupled regions.

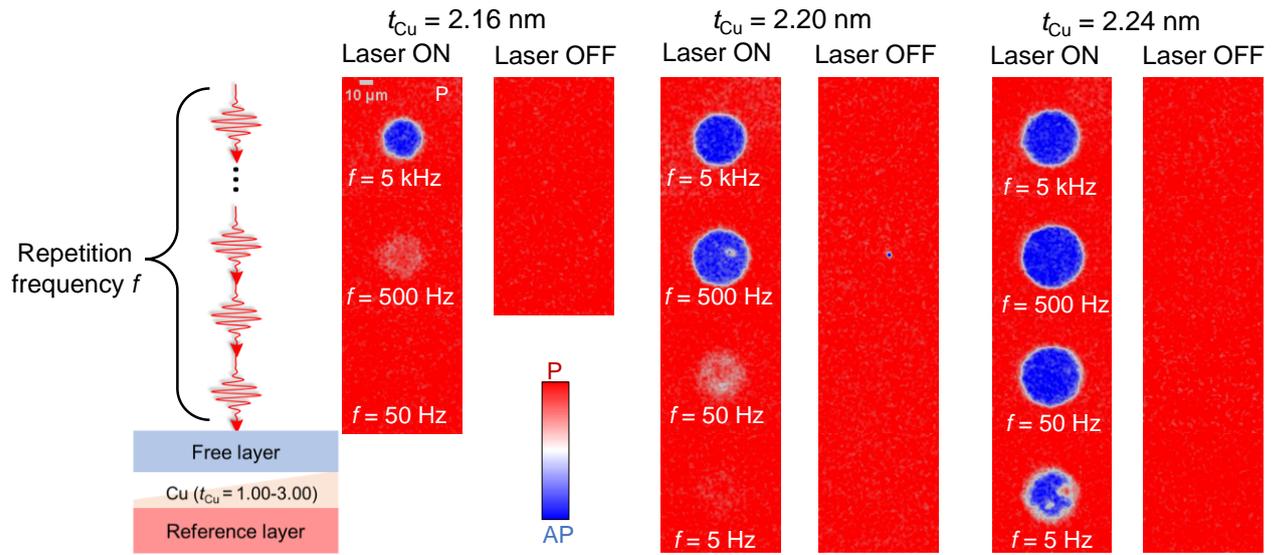

FIG. 4. The P-to-AP switching obtained by constant excitation with various frequency $f$. MOKE images obtained during constant excitation (Laser ON) and after the excitation (Laser OFF). The laser fluence used for the experiment is around 7.8 mJ/cm$^2$.

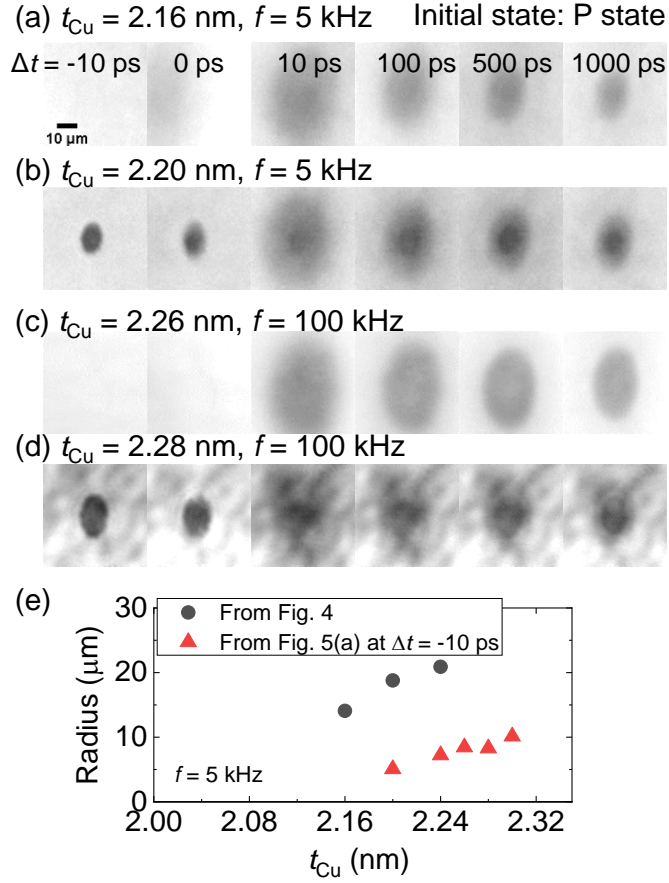

FIG. 5. MOKE images at various time delay Δ*t* obtained by TR-MOKE imaging method for different Cu thickness ($t_{Cu}$) and frequency (*f*). No external magnetic field is applied during the experiment. (a) $t_{Cu}$ = 2.16 nm, *f* = 5 kHz (b) $t_{Cu}$ = 2.20 nm, *f* = 5 kHz (c) $t_{Cu}$ = 2.26 nm, *f* = 100 kHz (d) $t_{Cu}$ = 2.28 nm, *f* = 100 kHz. The laser fluence used for the experiment is around 4.5 mJ/cm$^2$. (e) Radius (horizontal direction) of the reversed domains obtained from static (figure 4 for 5 kHz) and dynamic experiments at Δt = -10 ps as a function of $t_{Cu}$.

Supplementary Material for
# Influence of interlayer exchange coupling on ultrafast laser-induced magnetization reversal in ferromagnetic spin valves

Junta Igarashi[1,*], Yann Le Guen[1], Julius Hohlfeld[1], Stéphane Mangin[1,2], Jon Gorchon[1], Michel Hehn[1,2], and Grégory Malinowski[1]

[1] Université de Lorraine, CNRS, IJL, F-54000 Nancy, France

[2] Center for Science and Innovation in Spintronics, Tohoku University, 2-1-1 Katahira, Aoba-ku, Sendai 980-8577 Japan

[*] junta.igarashi@univ-lorraine.fr, jigarashi1993@gmail.com

**Note 1 MOKE hysteresis loops obtained by applying a stronger magnetic field**

Figure S1 shows MOKE hysteresis loops obtained by applying an external field up to 700 mT perpendicular to the film plane in the thin Cu region. As single-step hysteresis loops are observed from $t_{Cu}$ = 1.00 nm to 2.12 nm, we conclude that the free and reference layers are strongly ferromagnetically coupled.

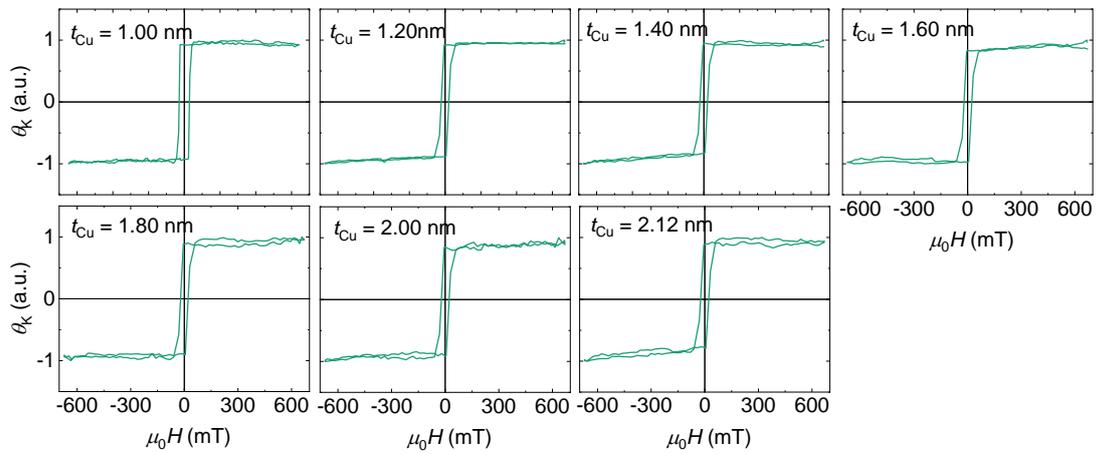

FIG S1 Normalised Kerr angle as a function of the applied magnetic field applied perpendicular to the sample plane for various Cu thickness ($t_{Cu}$).

**Note 2 Analysis of MOKE images to extract $F_{th}$**

Here, we demonstrate the analysis of MOKE images to extract $F_{th}$. Line profiles along horizontal and vertical directions were obtained, as shown in Fig. S1(a) and (b). The domain diameter for each state was extracted along both directions from the differential line profiles by detecting each peak (Fig. S1(b)). Subsequently, we calculated the domain size using the obtained domain diameter. The following equation was employed to fit the results presented in Fig. S2(c) and extract the $F_{th}$ of each state as a fitting parameter: Domain size = $\pi s w_0^2 [\ln\{E/(F_{th}\pi s w_0^2)\}]$, where $s$ is the aspect ratio between both directions determined from MOKE images (~1 for studied samples) [1].

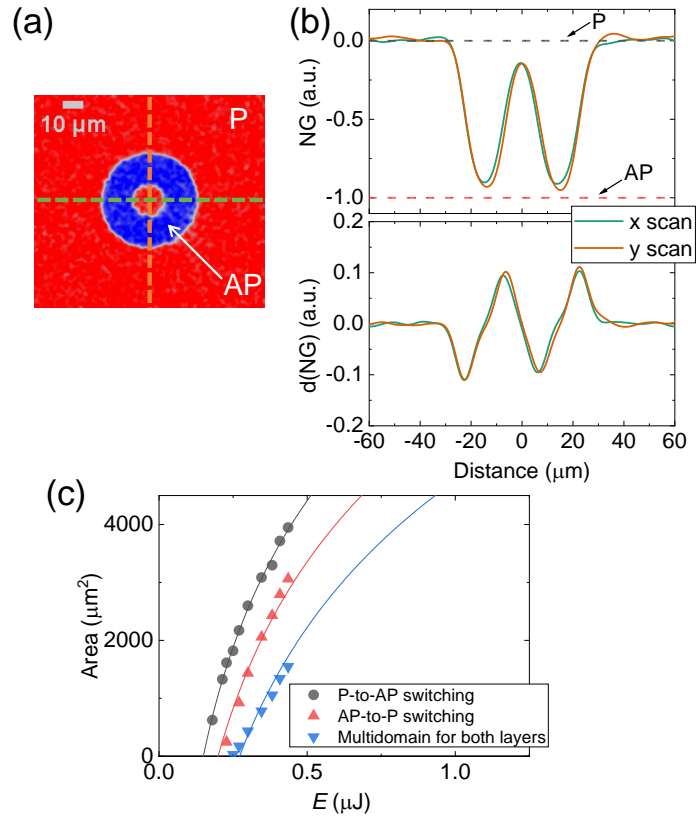

FIG S2  (a) MOKE image obtained after irradiation of a laser pulse starting from the P state (red) for a copper thickness $t_{Cu}$ = 2.32 nm and a laser fluence $F$ ~ 5.9 mJ/cm². The laser pulse is shined from the reference layer. (b)

Normalized gray level (NG) and the differential normalized gray level (d(NG)) as a function of the distance from the center of the laser (c) Domain area extracted as a function of the laser pulse threshold energy ($E$) for P to AP switching (gray circles), AP to P switching (Red triangles) and multidomain state in both layers(blue triangles). The curves represent fitting results using the equation Domain size = $\pi s w_0^2 [\ln\{E/(F_{th}\pi s w_0^2)\}]$.

**Note 3 Results obtained by shining the laser pulse from the reference layer**

Here, we present results obtained by shining the laser pulse from the reference layer. Figure S3 shows representative results after exciting the sample with a single laser pulse with a fluence of around 5.6 mJ/cm$^2$. From Fig. 2(a) in the main text and Fig. S3, we confirm no substantial difference depending on the direction of laser pulse excitation. In Fig. S4, we summarize the evolution of the threshold fluences ($F_{th}$) for the P-to-AP switching (black symbols), AP-to-P switching (red symbols), and multidomain formation for both layers (blue symbols). As for the results with irradiation of a laser pulse from the free layer, we added some data measured with samples from the previous study [2] (Fig. S4(b)). We also observed the same trend with the laser shining on the layer, as shown in the main text.

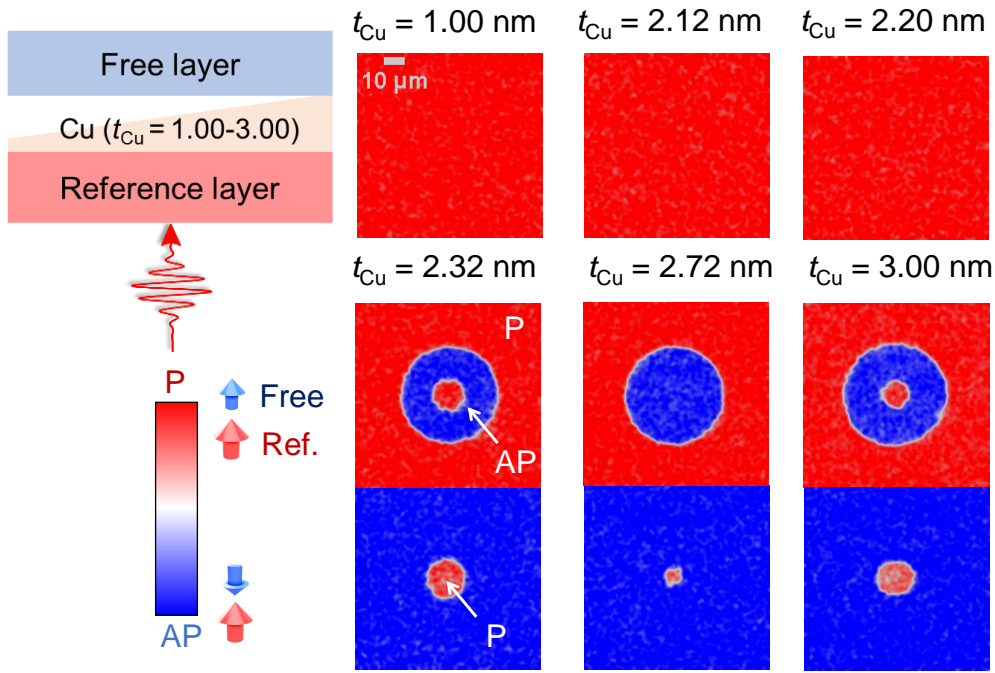

FIG S3 Results of magnetization switching using a single femtosecond laser pulse excited from the reference layer side. MOKE images obtained after irradiation of a laser pulse starting from the P state (red) and AP state (blue) for different Cu thickness ($t_{Cu}$).

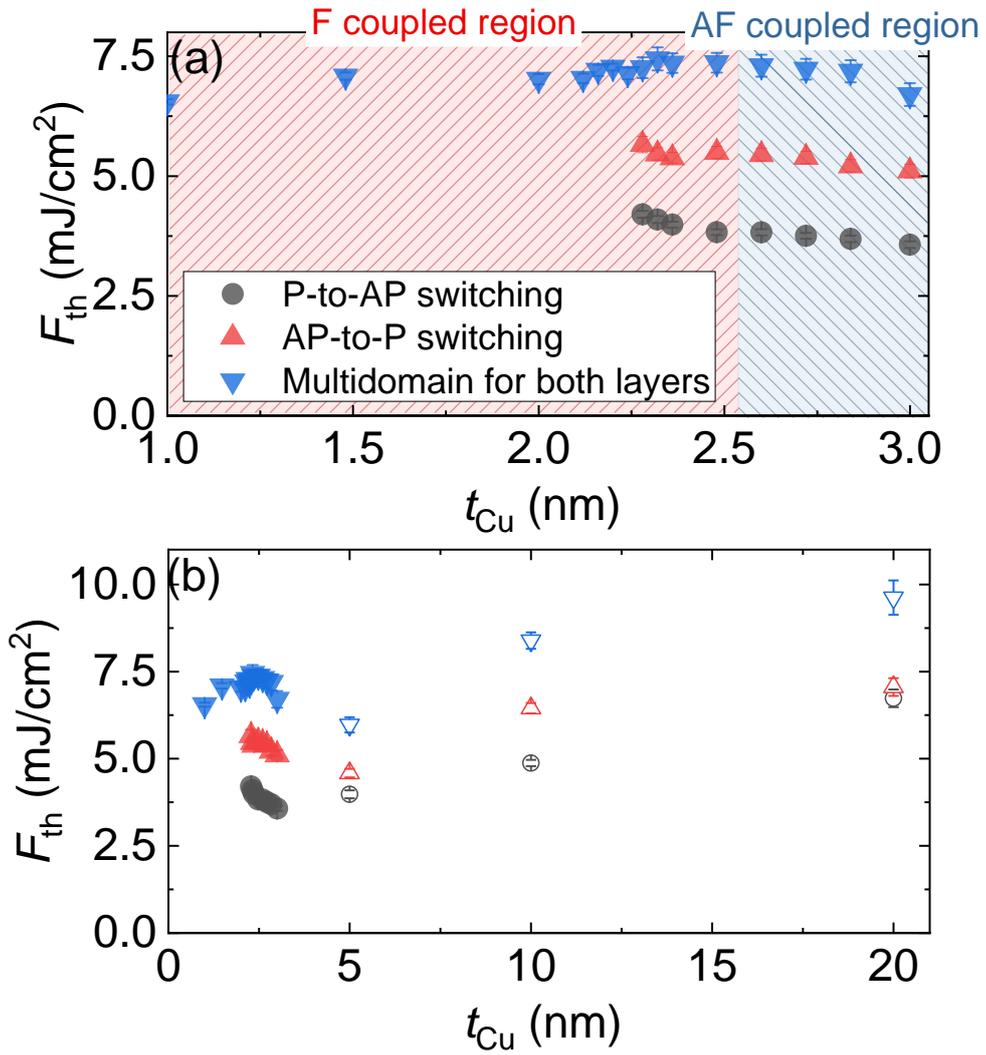

FIG S4 Threshold fluence for P to AP switching (gray circles), AP to P switching (red triangles) and multidomain state in both layers (blue triangles) as a function of Cu spacer thickness for (a) $t_{Cu}$ up to 3.0 nm. (b) $t_{Cu}$ up to 20 nm, including samples from a previous study (opened symbols) [2]. Shaded regions in (a) indicate ferromagnetic (F) coupled and antiferromagnetic (AF) coupled regions.

**Note 4 Comparing contrast level between MOKE image and Hysteresis loop**

Here, we discuss the magnetic state of the domain obtained by time-reloved MOKE imaging experiment. Figure S5 displays the MOKE hysteresis loop, the MOKE image obtained at a time delay of -10 ps, and its line profile along horizontal directions at $t_{Cu}$ = 2.20 nm and $f$ = 5 kHz. We used the same acquisition parameters of the camera during for MOKE hysteresis loops and TR-MOKE imaging experiments. We obtained MOKE hysteresis loop by extracting the mean gray level from the region of interest (ROI) covering the area of the reversed domain. The setup used for the time-resolved MOKE imaging and the hysteresis loop in figure S5(a), make use of polarizers in a near crossed configuration. If the magnetically induced rotations lie on both side of the fully crossed case, the resulting contrast levels will be asymmetric as the intensity will fold back around its minimum. In the case of this experiment, the position of the analyzer was chosen to optimize the contrast between the P+ and AP+ states. It just so happens that a way to get the best contrast between those two levels is to fold over the P- level in such a way that the P- and AP- levels are crushed near the minimum intensity leaving a bigger range for the other two contrast levels. The gray level between the P and AP states is around 1250, similar to that obtained from Fig. S5(b). Hence, the nucleated domain (Fig. S5(b)) is identified as a reversed domain at AP state.

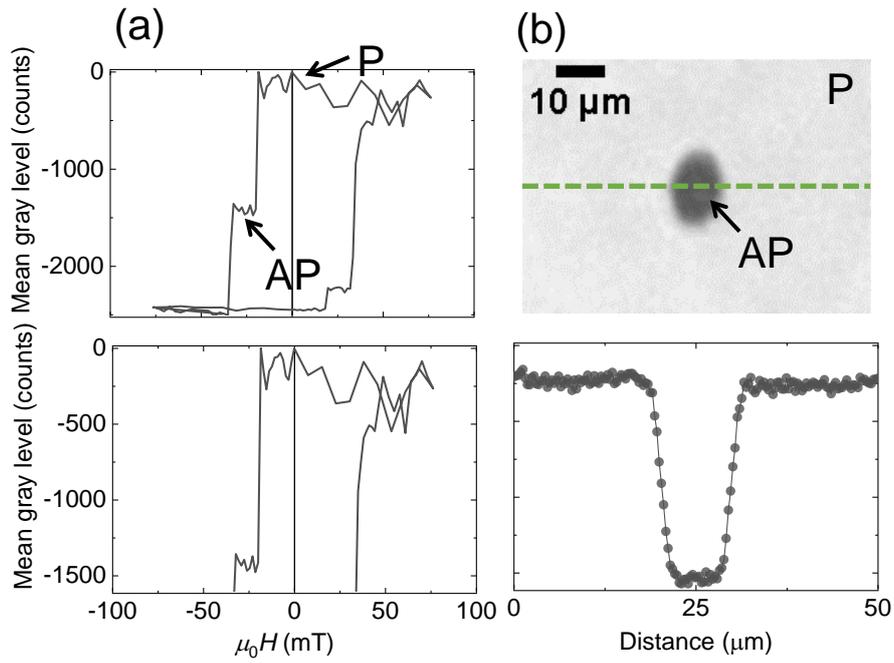

FIG S5 (a) MOKE hysteresis loop at $t_{Cu}$ = 2.20 nm obtained with the same setup for TR-MOKE imaging. Enlarged graph at bottom left. We plot the gray level obtained by subtracting the value at the P state and zero magnetic field. (b) MOKE image at a time delay of -10 ps. The image is created by subtracting the background image (P state without pump) from the original one. Its line profile of along the horizontal direction (green broken line in (b)).